\documentclass[aps,physrev,preprint,groupedaddress]{revtex4-2}

\usepackage{graphicx}
\usepackage{dcolumn}
\usepackage{bm}
\usepackage{amsmath}
\begin{document}


\title{Vertical microcavities with optical Kerr materials}


\author{Riya Varghese}
\email[]{riya.varghese@tuni.fi}
\author{Ali Panahpour} 
\email[]{ali.panahpour@tuni.fi}
\author{Marco Ornigotti} 
\email[]{marco.ornigotti@tuni.fi}
\author{Mikko J Huttunen}
\email[]{mikko.huttunen@tuni.fi}
\affiliation{Photonics Laboratory, Physics Unit, Tampere University, Korkeakoulunkatu 3, 33720 Tampere, Finland}


\date{\today}

\begin{abstract}
Optical Kerr nonlinearity is central to optical bistability in high-$Q$ microcavities. Here, we show that optical Kerr nonlinearities in Fabry--Pérot resonators can also result in self-focusing effects, that can substantially modify the $Q$-factors of the resonators. For the studied cavities, up to five-fold increase in $Q$-factors is predicted, attributed to reduction of sidewall leakage, confirmed via simulations performed on curved cavities. We also show that for high $Q$-factor cavities, the higher-order Kerr effect can become relevant already at modest input intensities scaling inversely with $Q$-factor for a particular $n_2/n_4$ ratio. We investigate numerically the role of higher-order Kerr effect on optical bistability responses and find that its inclusion can shift the bistability threshold to higher input intensities than predicted by the standard $n_2$-only model.
\end{abstract}


\maketitle

\section{Introduction}
\noindent High quality factor ($Q$-factor) microcavities have emerged as an important platform for enhancing occurring light--matter interactions \cite{AKavokin2017Microcavities}. Advances in fabrication technology have enabled a variety of high-$Q$ resonator architectures including whispering gallery mode microresonators \cite{Feng2012SiliconPerspective, Vernooy1998High-QInfrared}, microtoroids \cite{Armani2004ElectricalResonators}, and photonic crystal cavities \cite{Englund2005GeneralCavities} which can reach up to $Q$-factors of $10^{9}$ \cite{DelHaye2013Laser-machinedOptics, Armani2003Ultra-high-QChip, Soltani2007Ultra-highPhotonics}. The associated large field enhancement and changes in the photonic density of states in these structures \cite{Panahpour2026PhotonicMicrocavities}, can dramatically modify the efficiency of light–matter interactions, enabling applications across diverse areas such as low-threshold lasers \cite{Rong2007Low-thresholdLaser,Loncar2002Low-thresholdLaser,Dimopoulos2022Electrically-DrivenThreshold}, Purcell-enhancement of single-photon emission \cite{Panahpour2025PurcellMicrocavities,Engel2023PurcellMicrocavity}, nonlinear frequency conversion \cite{Lin2017NonlinearResonators, Wang2021High-QPhotonics} and all-optical switching \cite{Chai2017UltrafastSwitching}.

High $Q$-factor microcavities have also attracted interest for enhancing occurring nonlinear light--matter interaction \cite{Waks2006DispersiveCavity, Lu2019EfficientNanophotonics} that strongly benefits of the high field enhancement of the associated cavity field, as for example, second-harmonic generation in photonic crystal nanocavities \cite{Rivoire2009SecondPower}, Kerr-frequency-comb generation in high-$Q$ microring resonators \cite{DelHaye2007OpticalMicroresonator} and soliton formation in microresonators \cite{Moille2024ParametricallyMicrocavity}. Among these, optical Kerr nonlinearities \cite{Boyd2017NonlinearOptics} are of significant technological interest since they can enable ultrafast pulse generation by Kerr-lens modelocking \cite{Keller2003RecentLasers}, Kerr-induced frequency-comb generation \cite{DelHaye2007OpticalMicroresonator} and all-optical switching by optical bistability \cite{ Acklin1993BistableReflector}. The latter, in particular, refers to a regime where a device can exhibit two stable output states for the same input intensity \cite{Min2007OpticalMaterial} and it is especially attractive for photonic technologies because it enables optical memory \cite{Almeida2004OpticalChip} and all-optical switching functionalities \cite{Sankey1992AllopticalStructure}.

Interestingly, the optical intensities inside high-$Q$ cavities can reach very high levels (${\sim}\,$GW/cm$^2$) already with modest input intensities \cite{DelHaye2007OpticalMicroresonator,Vahala2003OpticalMicrocavities, Konthasinghe2017Self-sustainedMicrocavities}. In this regime, optical Kerr effect can significantly modify the cavity dynamics. First, it can cause lensing effects by Kerr self-focusing \cite{Chiao1964Self-TrappingBeams}, potentially affecting the cavity properties of a vertical Fabry--Pérot resonator supporting multiple lateral modes \cite{Yanagimoto2025DesignOscillator}. Second, the combined effect of intensity-dependent refractive index and feedback provided via the cavity mirrors can make the system optically bistable. Third, at sufficiently high intensities, the conventional perturbative description of nonlinear optics may no longer accurately describe phenomena such as saturation of the nonlinear refractive index \cite{Reshef2017BeyondMaterials} resulting in qualitatively incorrect predictions for resonance shifts and bistability thresholds \cite{Huttunen2020TransientMaterials}. Although Kerr-induced optical bistability in resonant cavities has been extensively studied \cite{Priem2005OpticalStructures., Acklin1993BistableReflector}, it remains an area of active research specifically in Bragg cavities. Recent work has demonstrated optical bistable switching in photonic-crystal cavities \cite{Pradhan2025EnhancedGeneration} and low-power bistable switching in chirped Bragg gratings \cite{Sudhakar2022Low-powerMixing}.

Here, we first investigate Kerr self-focusing from the $n_2$ nonlinearity and its effect on the $Q$-factor of a finite Bragg cavity. As the input intensity increases, the Kerr-induced lensing can start improving the $Q$-factor by reducing sidewall leakage. Here, we report nearly five-fold enhancement of the $Q$ values, becoming comparable to those of geometrically curved Bragg cavities. We then examine how the higher-order Kerr effect (HOKE) term \cite{Loriot2009MeasurementComponents, Borchers2012SaturationSolids} such as $n_4$ modify the optical bistability for the case of a very high-$Q$ cavity. The input intensities at which the higher-order term becomes relevant depend on the the inverse of the cavity $Q$-factor, as well on the strength of higher-order term ($n_2/n_4$ ratio). For a Bragg cavity with $Q$-factor of $10^4$ and $n_2/n_4$ ratio of $-1\times10^{13}$, the HOKE term can become non-negligible already at modest input intensities of a few MW/cm$^2$. In this regime, the optical bistability response departs from the prediction of the $n_2$-only Kerr model, resulting in a shift of the bistability threshold to higher intensities. We further extend the study to account for the full spatial distribution of the intracavity field throughout the resonator. This results in a larger resonance shift and a lower bistability threshold than predicted by the core-only approximation, with the transmittance differing by up to 17.5 $\%$ at the highest input intensity considered.

Our paper is organized as follows: In Sect.~\ref{sec:theory}, the temporal coupled-mode theory was used to describe the resonance shifts due to HOKE. All-dielectric cavity design and optical bistability analysis framework is described in Sec.~\ref{sec:MandM}. Section~\ref{sec:RandD} looks into the effects of Kerr self-focusing on the cavity $Q$-factor, investigates the threshold at which higher-order Kerr effects become significant in high-$Q$ cavities, and analyzes their impact on the optical bistability response. Finally, Section~\ref{sec:Concl} provides the conclusions.

\section{\label{sec:theory}Theory}
\noindent The dynamics of an optical bistable system can be described using the temporal coupled-mode theory \cite{Haus1984WavesOptoelectronics}:
\begin{equation}
\frac{da}{dt} = j \omega_0 a - \frac{1}{{\tau_0} + {\tau_e}} a + \sqrt{\frac{2}{\tau_e}}\, s_{in}\,,
\label{eq:cmt}
\end{equation}
where $a$ is the cavity mode amplitude, $\tau_0$ is the cavity decay rate, $\tau_e$ is the decay rate due to external coupling to the cavity, $s_{in}$ is the amplitude of the input wave traveling into the cavity and $\omega_0$ is the resonant frequency of the cavity given by 
\begin{equation}
\omega_0 = \frac{\pi c}{n_0l}\,,
\label{eq:omega}
\end{equation}
where $l$ is the length of the cavity, $c$ is the speed of light and $n_0$ is the refractive index. In a high-$Q$ cavity, the large intracavity intensity gives rise to the optical Kerr effect, a nonlinear phenomenon in which the refractive index becomes dependent on the optical intensity, thereby modifying the optical response of the cavity. As the input intensity increases, higher-order nonlinear terms, such as $n_4$, will need to be considered due to the onset of moving towards non-perturbative regime \cite{Reshef2017BeyondMaterials}. For simplicity, we consider the refractive index of a non-magnetic and centrosymmetric material 
\begin{equation}
n = n_0 + n_2 I + n_4 I^2 + ...\,,
\label{eq:HOKE terms}
\end{equation}
where $n_2$ and $n_4$ are the nonlinear coefficients corresponding to third- and fifth- order susceptibilities given by $n_2 = 3\chi^{(3)}/4 n_0^2 \varepsilon_0 c$ and $n_4 = 5\chi^{(5)}/8 n_0^3 \varepsilon_0^2 c^2$. The intracavity intensity is related to the mode amplitude $a$ by 
\begin{equation}
I = \frac{c|a|^2}{Aln_0}\,,
\label{eq:intensity}
\end{equation} 
where $A$ is the effective mode area and $l$ is the cavity length. Substituting the above equation into the intensity-dependent refractive index in Eq.\eqref{eq:HOKE terms} will shift the resonant frequency from $\omega_0$ to
\begin{equation}
\omega = \omega_0 \left[1 - \Bigg|\frac{a}{a_0}\Bigg|^2 + \Bigg|\frac{a}{a_0}\Bigg|^4 \left(1-\frac{n_4 n_0}{n_2^2}\right)\right]\,,
\label{eq:new omega}
\end{equation}
where $|a_0|^2 = n_0^2 A l/n_2 c$. The cavity redshifts as a result of the linear $n_2$ term and as the intracavity intensity further increases, the higher-order term $|a/a_0|^4$ becomes important. This leads to the saturation or enhancement of the redshift depending on the sign of $n_4 n_0/n_2^2$ term. For $n_4 n_0/n_2^2\;<\;0$, the overall coefficient becomes positive and counteracts the redshift \cite{Chen2006MeasurementGlasses} while a positively-valued term reinforces the redshift \cite{Tarazkar2014High-orderAtoms}. Here, we focus on the first case where the $n_4 n_0/n_2^2\;<\;0$. At low intensities, the linear Kerr term dominates the responses, while at higher intensities $n_4$ becomes non-negligible and comparable to the $n_2$ contribution. Interestingly, this optically-induced cavity resonance detuning gives rise to optical bistability, making it possible for the cavity to start supporting two different cavity modes \cite{Tamm1990BistabilityLaser}. The $Q$-factor of the cavity plays an important role in reaching this regime. Since the HOKE depends on the fourth power of the amplitude of the cavity field, such contributions could start taking place at very modest input intensity levels, when compared against studies on bulk materials \cite{Lenz2000LargeGlasses, Borchers2012SaturationSolids}.

To quantify the departure from linearity of the Kerr-type response, we define the refractive index change $\Delta n_{\text{lin}}(I)$ obtained from only the $n_2$ term and, from Eq.~\eqref{eq:HOKE terms} we define $\Delta n_{\text{HOKE}}(I)$ as the refractive index change including the higher-order term proportional to $n_4$. The relative deviation between the two refractive index changes is given by
\begin{equation}
\Delta(I) = \frac{(\Delta n_{\text{HOKE}}(I)-\Delta n_{\text{lin}}(I))}{\Delta n_{\text{lin}}(I)} = \frac{n_4 I}{n_2}.
\label{eq:deviation}
\end{equation}
This ratio helps to quantify the relative contribution of $n_4$ to $n_2$ at a given input intensity. The threshold intensity at which $\Delta(I)$ exceeds 5$\%$ is denoted as $\Delta_{5\%}$.

\subsection{\label{sec:MandM}Materials and Methods}
\begin{figure}
\includegraphics{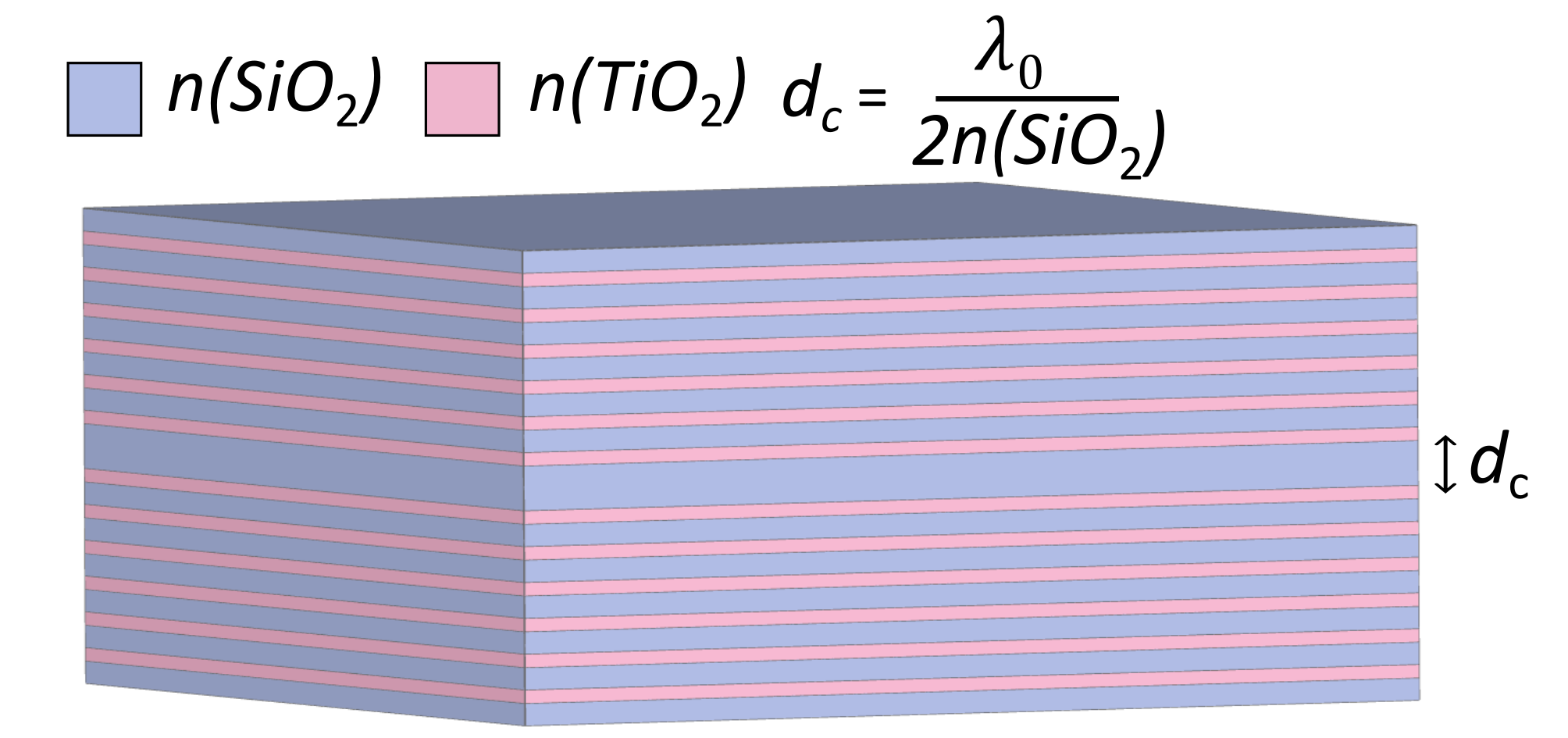}
\caption{The schematic of the all-dielectric Bragg cavity considered.}
\label{fig:Fig0}
\end{figure}

\noindent We consider all-dielectric Bragg cavities as shown in Fig. \ref{fig:Fig0}, composed of fused silica (SiO$_2$) and titanium dioxide (TiO$_2$) for our study. The resonant cavity is designed to operate at the wavelength $\lambda_0$ = 1550 nm. Although SiO$_2$ has lower nonlinear refractive index compared to other nonlinear materials, it is chosen as the core of the cavity because of its wide optical transparency, negligible absorption at the operating wavelengths, and high damage threshold (in ranges of 100 GW/cm$^2$ \cite{Chimier2011DamageMechanisms}), which together could enable stable operation at high intensities. 

The optical response of the Bragg cavity is modeled using the transfer-matrix method \cite{Yeh1988OpticalMedia}, which allows direct calculation of the electric-field distribution across the multilayer structure for a given excitation wavelength and input intensity. Due to the strong field confinement at the cavity resonance, the electric field is maximized at the core layer of the Bragg structure (Fig.~\ref{fig:Fig1}a). Consequently, the nonlinear refractive index is included only in the core layer (SiO$_2$ core), as the nonlinear contribution from the surrounding layers is negligible compared to that of the cavity region. We take $n_2$ = $3 \times 10^{-16}\,\mathrm{cm}^2/\mathrm{W}$ and $n_4$ = $-1 \times 10^{-30}\,\mathrm{cm}^4/\mathrm{W}^2$ from literature ($n_2/n_4$ ratio = $-3\times10^{14}$) \cite{Ekvall2001StudiesSilica}. The bistability curve was obtained by introducing an external coupler with an input field $s_{in}$ and with a linewidth greater than the intrinsic cavity linewidth so that the transmission is maximized at resonance. The detailed methodology of the curved cavity designs and the numerical procedure for the optical bistability calculations is provided in the Supplementary Material \cite{SM}.

\section{\label{sec:RandD}Results and Discussions}
\begin{figure*}
\includegraphics[width=\textwidth]{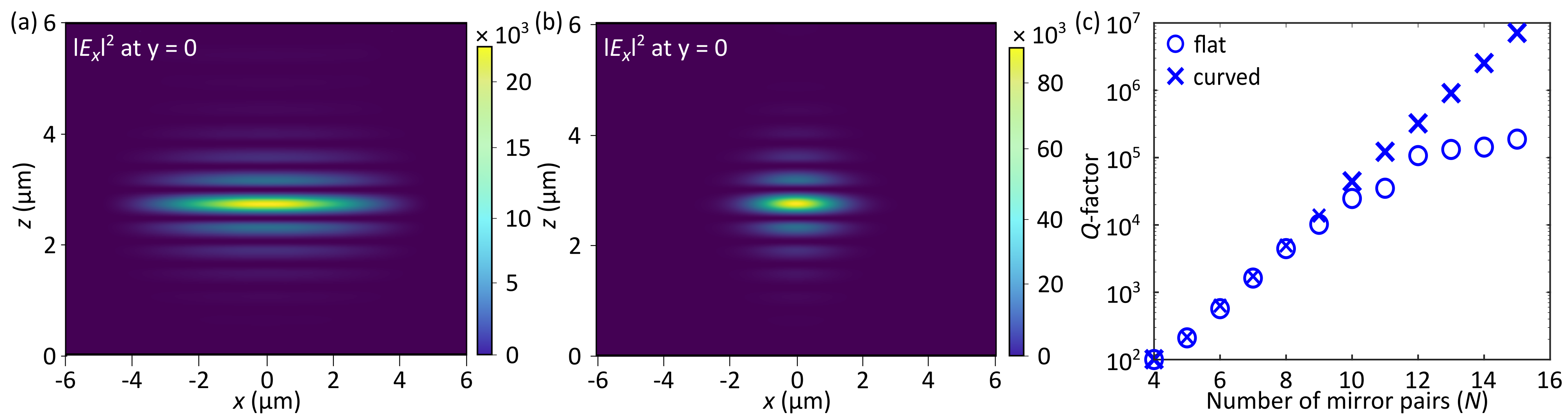}
\caption{a) The intracavity field inside a cavity with flat DBR mirror ($N=6$) on top and bottom b) The intracavity field inside a cavity with curved DBR mirror on top and flat DBR mirror on bottom ($N=6$) with a gaussian excitation in both cases. c) The comparison of $Q$-factor for cavity with flat mirror on top and a curved mirror on top.}
\label{fig:Fig1}
\end{figure*}

\noindent While planar and laterally infinite Bragg cavities provide a convenient framework for studying cavity modes, they deviate from experimentally relevant cavity configurations where lateral confinement plays a crucial role. To understand the influence of lateral confinement on mode localization and $Q$-factor, we first investigate finite flat Bragg cavities and compare their modal properties with those of curved Bragg cavities as a reference. We consider a cavity with a curved top Bragg mirror and a flat bottom Bragg mirror with six mirror pairs each. In Fig.~\ref{fig:Fig1}a, b we compare the field distributions of flat and curved mirror cavities excited by a Gaussian beam mode matched to the fundamental cavity mode, with its waist positioned at the cavity center. The field in the planar cavity extends over a much larger transverse region  compared to the case of curved cavity (Fig.~\ref{fig:Fig1}a), resulting in weaker confinement and lower peak intensity ($\left| E_x \right|^2\;=\;2 \times 10^4$). In stark contrast, the optical field inside the curved cavity is tightly confined around the cavity center (Fig.~\ref{fig:Fig1}b) and reaches an intracavity intensity approximately four times larger than that obtained inside the planar cavity. We attribute the stronger confinement to be due to the focusing effect of the curved mirror which suppresses lateral spreading. 

We also compare the evolution of the $Q$-factor as a function of the number of Bragg mirror pairs ($N$), as shown in Fig.~\ref{fig:Fig1}c. For cavities containing approximately 4--8 mirror pairs, the $Q$-factors of the flat and curved geometries remain nearly identical. In this regime, the dominant loss mechanism is transmission through the mirrors, and therefore the cavity geometry has little influence on the overall losses. However, as the mirror reflectivity increases with additional layer pairs, the lateral losses through the sidewalls begin to dominate. As a result, when the number of mirror pairs increases above $N=8$, the $Q$-factors of the two cavities start visibly deviating from each other. While the $Q$-factors of the flat cavities continue to increase, the curved cavities exhibit a substantially more rapid increase of the $Q$-factor values. Consequently, the difference in $Q$-factors between the two geometries becomes increasingly pronounced 
when further increasing the number of mirror pairs $N$. At $N=15$ mirror pairs, the curved cavity reaches a $Q$-factor of $7 \times 10^6$ compared 
against the value of only $2 \times 10^5$ for the flat cavity, corresponding to approximately 35-fold increase in the $Q$ value. These results demonstrate that geometric curvature is an effective way to enhance mode confinement and $Q$-factor in finite Bragg cavities. We now evaluate whether an equivalent enhancement can be achieved in a flat cavity through optical Kerr nonlinearity alone.  
\begin{figure}
\includegraphics{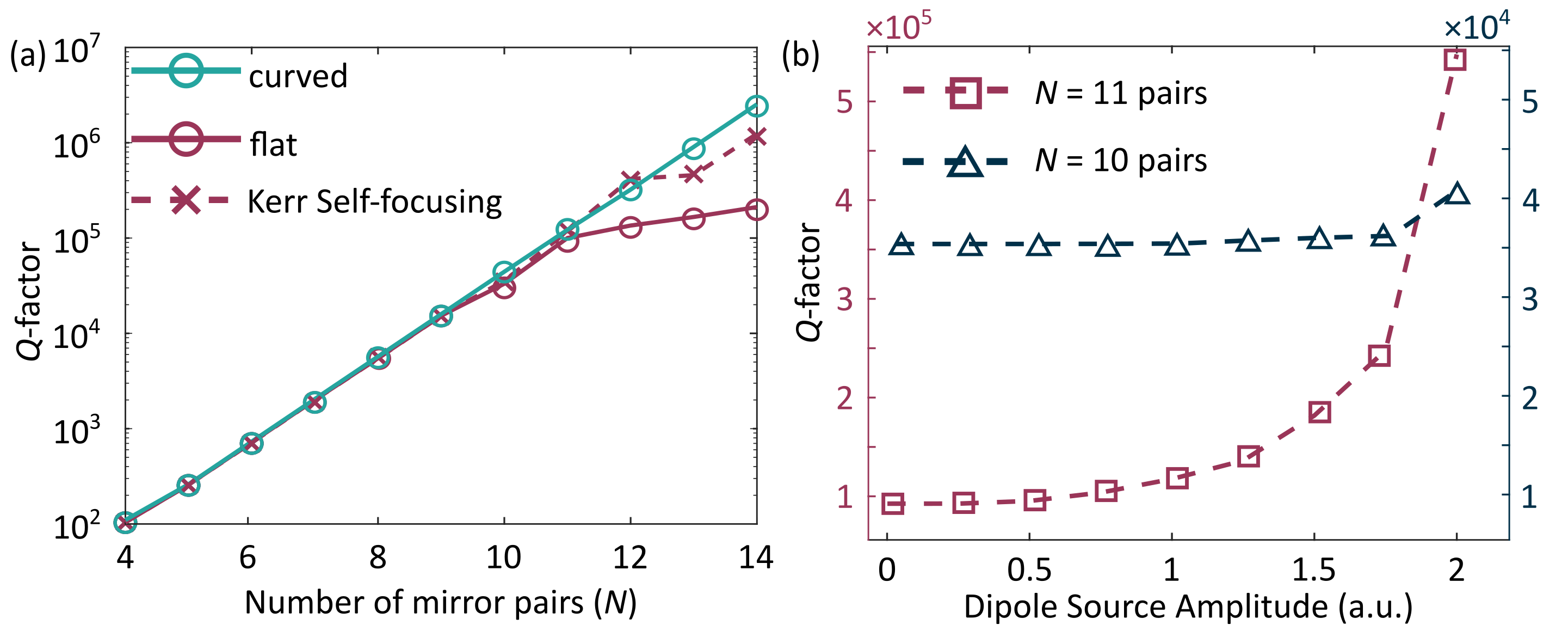}
\caption{a) Comparison of calculated $Q$-factors as functions of mirror pair number $N$ for curved and flat cavities. The $Q$-factors of flat cavities when Kerr effect is considered is also shown (cross marks) and is comparable to those curved cavities. b) The increase in $Q$-factor as a function of dipole source amplitude, placed within the center of the core region. Increase of source amplitude gives rise to Kerr lensing, resulting in increased $Q$-factors. The effect becomes more pronounced when the number of mirror pairs $N$ is increased, emphasizing the role of sidewall leakage in the overall cavity $Q$-factor.}
\label{fig:Fig2}
\end{figure} 

Next we introduce an intensity-dependent refractive index into the core of the finite flat cavity by introducing a Kerr nonlinear model in the simulation domain. The Fig.~\ref{fig:Fig2}a compares the $Q$-factors of the flat Bragg cavities without Kerr effect (pink circle markers) and with a Kerr nonlinear refractive index (pink cross markers). The cavities with Kerr effect have higher $Q$-factors that are comparable to those of the curved cavities. For example, Kerr lensing results in a 5-fold increase of the $Q$ value for a flat cavity consisting of $N=14$ pairs. We investigated the enhancement of $Q$-factor with the amplitude of a dipole excitation source for different mirror pairs (Fig.~\ref{fig:Fig2}b). The $Q$-factor calculation procedure is given in Supplementary Material \cite{SM}. As the input amplitude increases, the $Q$-factor rises for both $N=10$ and $N=11$ pairs, with $N=11$ showing an approximately 500$\%$ increase and $N=10$ an approximately 15$\%$ increase relative to their respective low-amplitude values. Although the cavity resonance red-shifts due to the Kerr effect, the resulting self-focusing enhances the $Q$-factor of flat cavities, bringing their performance closer to those of geometrically curved cavities.   

\begin{figure*}
\includegraphics[width=\textwidth]{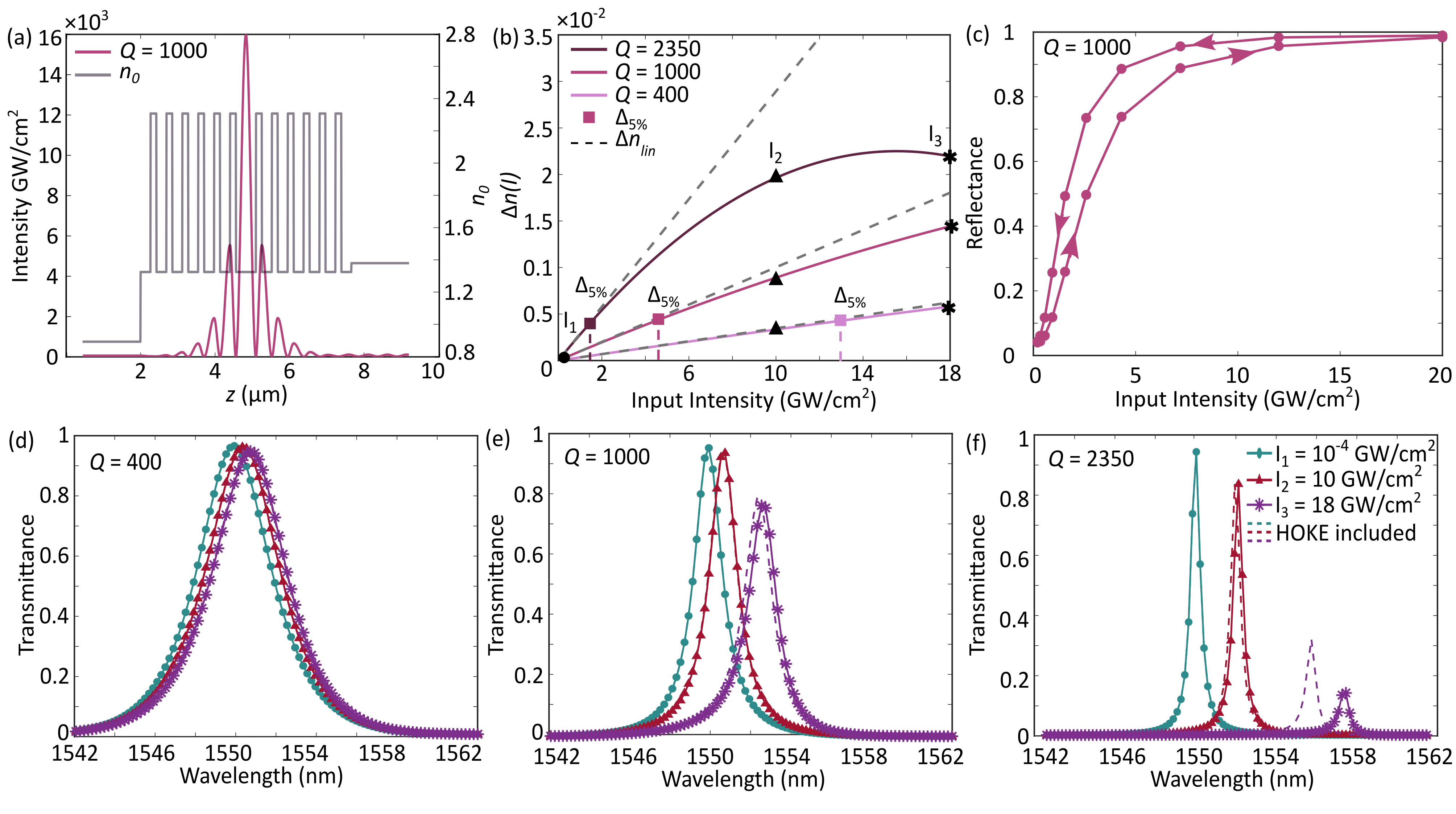}
\caption{a) The electromagnetic field inside the cavity at input intensity of 10 GW/cm$^2$ for a Bragg cavity of $Q$-factor = 1000. b) $\Delta n_{\text{lin}}(I)$ (dashed lines) and $\Delta n_{\text{HOKE}}(I)$ (solid lines) due to Kerr effect for cavities of varying $Q$-factors. $\Delta_{5\%}$ is marked in square markers. c) The hysteresis curve showing the optical bistability for a Bragg cavity of $Q$\;=\;1000. Arrows are used to indicate the direction of the variation of input intensity. The transmittance response for different input intensities (as marked in Fig.\ref{fig:Fig1}b, circle, triangle, asterisks markers) for a Bragg cavity of d) $Q$\;=\;400 e) $Q$\;=\;1000 and f) $Q$\;=\;2350 when only $n_2$ term is considered. The dashed lines represent the transmittance response when the HOKE term $n_4$ is also considered.}
\label{fig:Fig3}
\end{figure*}

The strong field enhancement achieved in these cavities can make higher-order nonlinearities increasingly relevant. In order to investigate the role of HOKE, we numerically analyse how the refractive index profiles and resonance peak positions are modified as functions of the input intensity and the $Q$-factor of the resonators. We start by considering moderate-$Q$ cavities ($Q$-factor = 400, 1000, 2350), where the $Q$-factor is varied by increasing the number of Bragg mirror pairs, while keeping all other parameters fixed. We first analyse the case where the $n_2$ is only considered for which the index change is linear with respect to the input intensity (Fig.~\ref{fig:Fig3}b, dashed black line). When we include the HOKE term $n_4$, at lower intensity values the refractive index change is approximately linear with respect to the input intensity. However, at higher input intensities, the index change starts visibly deviating from linear dependence on intensities, and starts showing saturation effects (Fig.~\ref{fig:Fig3}b, solid line). 

In Fig.~\ref{fig:Fig3}b, the 5$\%$ threshold intensity ($\Delta_{5\%}$)(Eq.~\eqref{eq:deviation}) is marked by square markers to clearly visualize the dominance of higher-order terms. The highest-$Q$ cavity ($Q$ = 2350) shows a significantly stronger deviation from linear Kerr behaviour, with the $\Delta_{5\%}$ reached at a comparatively low input intensity of 1.56 GW/cm$^2$. The moderate-$Q$ cavity ($Q$ = 1000) shows intermediate behaviour, deviating from linearity later than the highest-$Q$ case and reaching the $\Delta_{5\%}$ at 4.50 GW/cm$^2$. The lowest-$Q$ cavity ($Q$ = 400) maintains an approximately linear refractive index response over the broadest intensity range, with the $\Delta_{5\%}$ occurring only at a substantially higher input intensity of 12.97 GW/cm$^2$. The enhanced field confinement in higher-$Q$ cavities amplifies the effective intra-cavity intensity, thereby increasing the relative contribution of HOKE terms. As a result, the linear Kerr regime breaks down more rapidly as the cavity $Q$-factor increases especially for high-$Q$ resonators ($Q\;\approx\; 10^4$) and will be later discussed in detail. 

These nonlinear refractive index changes directly modify the cavity resonance condition, introducing a feedback mechanism between the intracavity field and the spectral response of the cavity leading to optical bistability. The optical bistability is characterised by two stable output values for a single input. This can be visualised by the hysteresis graph. We have plotted the hysteresis for a cavity with $Q$\;=\;1000 at 1550 nm (Fig.~\ref{fig:Fig3}c), which acted as the designed operation wavelength. While increasing the input intensity ($I_{in}$) from 0.1  GW/cm$^2$ to 20 GW/cm$^2$, the intensity-dependent reflectance follows a different path than when decreasing $I_{in}$ from  20 GW/cm$^2$ to 0.1  GW/cm$^2$. The combined optical Kerr effect and the feedback due to the cavity, tune the cavity into or out of resonance differently depending on the history of the intensity, giving rise to hysteresis.

Now in order to see the effect of HOKE in the optical bistability response, we plot the transmitted  intensity as a function of wavelength for selected input intensities. These intensities ($I_{in}$\;=\;$10^{-4}$ GW/cm$^2$, $10$ GW/cm$^2$, $18$ GW/cm$^2$) are indicated in Fig.~\ref{fig:Fig3}b as different markers (circle, triangle, asterisks) and chosen so that they are below and above the 5$\%$ deviation threshold for different $Q$-factors. The transmittance curve when $n_2$ is only considered is represented in solid curve and when HOKE term $n_4$ are introduced is represented as dashed lines in Fig.\ref{fig:Fig3} d,e,f. 

In Fig.~\ref{fig:Fig3}d, for a Bragg cavity with $Q$\;=\;400 at the lowest intensity of $10^{-4}$ GW/cm$^2$, the transmission spectra has a peak at 1550 nm which is the designed wavelength. When the input intensity is  increased to 10 GW/cm$^2$, the resonance wavelength visibly redshifts. When input intensity is  further increased to 50 GW/cm$^2$, the transmission further redshifts without significant drop in the transmittance. The effects of HOKE are not visible since the dashed curve overlaps with the solid lines. For a Bragg cavity with $Q$\;=\;1000 (Fig.~\ref{fig:Fig3}e), the same redshift can be observed as we increase the input intensities and is more prominent at the highest value input intensity of $I_{in}$\;=\;18 GW/cm$^2$ with a drop in the transmittance value. The effect of HOKE are also visible in the Fig.~\ref{fig:Fig3}e where the solid and the dashed curves don't overlap anymore at that intensity. Similarly for a Bragg cavity with $Q$\;=\;2350 (Fig.~\ref{fig:Fig3}f), as the input intensity increases the transmittance curve further redshifts and eventually the profile collapses from the Lorentzian shape. The effect of HOKE is even more prominent and the curve where higher-order terms are considered lags behind the one with only $n_2$ considered. At intensities where higher-order terms become relevant, the refractive index change saturates upon further increases in intensity, thus stopping the further redshift. As the $Q$-factor increases to $Q$\;=\;2350, the difference between the spectra with only $n_2$ and those including higher-order terms becomes more significant since the intracavity field enhancement inside high-$Q$ cavities is higher leading to prominent changes in the plot.  
\begin{figure*}
\includegraphics[width=\textwidth]{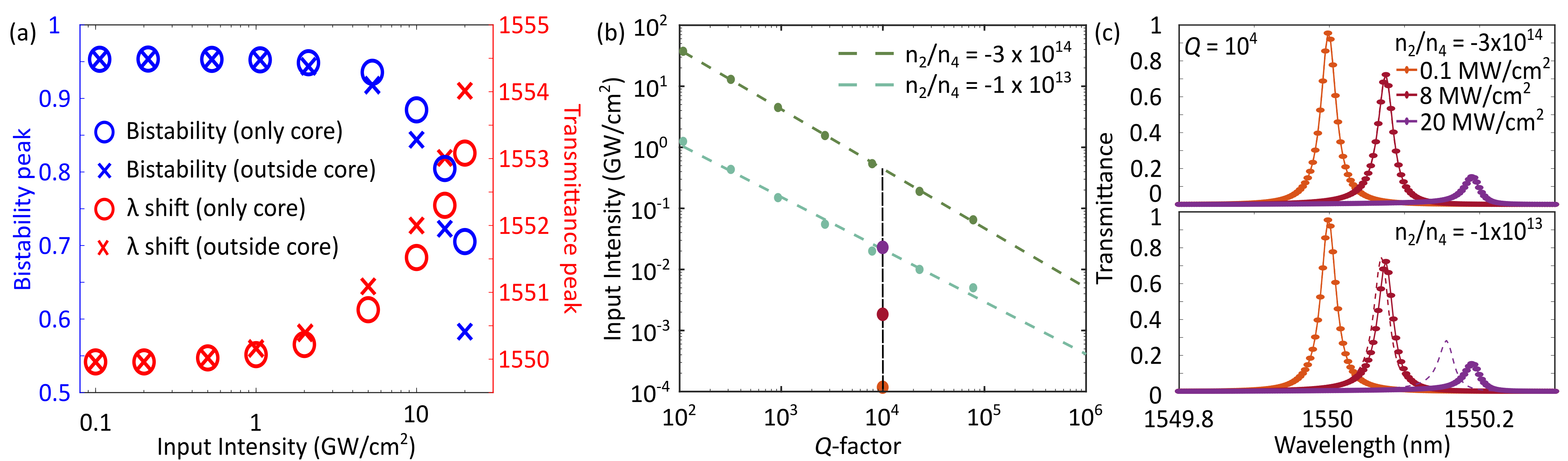}
\caption{a) The transmittance  peak shift and the optical bistability peak for a Bragg cavity with $Q$ = 1000 is shown  as a function of input intensities. The Kerr-induced refractive index change is shown for both the core only (in circles) and also outside the core (in cross). b) The 5$\%$ threshold ($\Delta_{5\%}$) intensity as a function of $Q$-factor for different higher-order Kerr contributions ($n_2/n_4$ ratio). c) The optical bistability curve for different $n_2/n_4$ ratio for different input intensities marked in Fig.~\ref{fig:Fig4}b.}
\label{fig:Fig4}
\end{figure*}

In realistic cavities, the optical field extends beyond the high-index core into the surrounding medium as seen in Fig.~\ref{fig:Fig3}a. Since the nonlinear refractive index change depends on the spatial distribution of intensity, neglecting the external field components leads to an underestimation of the effective nonlinear response. It is important to consider how differences in the intracavity field distribution can influence the transmission peak shift (Fig.~\ref{fig:Fig4}c, red markers) as a function of input intensity for both cases. We consider two cases where the intracavity field is assumed to be confined strictly within the core region (circles), and another where the full field distribution, including contributions outside the core, is taken into account (crosses) for a Bragg cavity of $Q$\;=\;1000. The results show that when only the core field is considered, the predicted resonance shift is smaller (Fig.~\ref{fig:Fig4}a, red circle markers) compared to the case where the extended field distribution is included (Fig.~\ref{fig:Fig4}a, red cross markers). This difference becomes more pronounced at higher input intensities, where nonlinear effects are stronger for the case where the field within the mirror pairs is also considered. Since the resonance peak shift is higher when the extended field outside the core is considered, this will directly affect the bistability. In Fig.~\ref{fig:Fig4}a, we plot the transmittance as a function of input intensities (blue markers) to see how the cavity becomes bistable as we increase the input intensity. The transmittance slowly drops, as seen in Fig.~\ref{fig:Fig3}e but the transmittance drops more quickly when the field outside the core is also considered. The difference in transmittance between the two models becomes increasingly significant at higher input intensities, with a difference close to 17.5$\%$ at 20 GW/cm$^2$ and only about 1$\%$ at 5 GW/cm$^2$ relative to the core-only model. This means that in realistic cases the bistability threshold will be even lower than predicted when the field contributions outside the core region are taken into account.

We now extend our studies to include high-$Q$-factor cavities to examine the input intensity at which the HOKE becomes relevant. We plot the input intensity (Fig.~\ref{fig:Fig4}b) where HOKE becomes relevant ($\Delta_{5\%}$) as a function of the $Q$-factor for different $n_2/n_4$ ratio. The threshold intensity depends inversely of the cavity $Q$-factor, with its slope determined by the $n_2/n_4$ ratio. Consequently, for materials with stronger higher-order nonlinear contributions ($n_2/n_4 = -1\times10^{13}$), HOKE becomes significant at lower input intensities, reaching values of only a few MW/cm$^2$ for sufficiently high $Q$-factors.

In order to examine the effect of HOKE on optical bistability response, we consider a cavity with $Q$ = $10^4$, operating at input intensities below and above the HOKE threshold (Fig.~\ref{fig:Fig4}c). To quantify this effect, the optical bistability curves were calculated using the same feedback configuration while varying only the nonlinear response of the medium. For $n_2/n_4$ = $-3\times10^{14}$, the selected input intensities ($I_{in}$ = 0.1 MW/cm$^2$, 8 MW/cm$^2$, 20 MW/cm$^2$) lie well below the HOKE threshold (Fig.~\ref{fig:Fig4}b, purple markers). As a result, the bistability curves are nearly identical (Fig.~\ref{fig:Fig4}c, top) to those predicted by the $n_2$ only Kerr model. However for $n_2/n_4$ = $-1\times10^{13}$) the onset of HOKE shifts to lower intensities. As a result, at an input intensity of 20 MW/cm$^2$ the bistability threshold when HOKE is considered (Fig.~\ref{fig:Fig4}c, below, dashed line) deviates from the prediction obtained using only the $n_2$ term (Fig.~\ref{fig:Fig4}c, below, solid line). Thus indicating that the conventional Kerr approximation overestimates the switching threshold when higher-order terms are neglected. Consequently, by considering the resonator $Q$-factor and nonlinear material parameters, the saturation effects due to higher-order Kerr effects can be minimized.

\section{\label{sec:Concl}Conclusions}
\noindent In this work, we find that Kerr-induced self-focusing of light within finite Bragg microcavities can result in increase of cavity $Q$-factors. The resulting response resembles that of a Bragg cavity with curved mirrors. We associate these effects to reduced sidewall leakage of the cavity modes. We analysed the influence of higher-order Kerr effect (HOKE) in the optical response and bistability of planar Bragg cavities numerically. At high input intensities HOKE becomes non-negligible and beyond this regime, the refractive index shift no longer follows the linear Kerr approximation, leading to modifications of the cavity resonance shift, transmittance response, and optical bistability.

In addition, we extended the analysis beyond the cavity core by incorporating the complete spatial field distribution throughout the Bragg structure. Accounting for the nonlinear response in the mirror regions leads to larger resonance shifts and more pronounced modifications of the bistability characteristics, particularly at higher input intensities. This highlights the importance of accounting for the spatially distributed field distribution when modeling realistic cavity systems.

We further examined the dependence of these effects on cavity $Q$-factor and showed that higher-$Q$ cavities reach the nonlinear regime at lower input intensities due to enhanced intracavity field buildup. The intensity at which HOKE contributions become significant was found to depend strongly on both the cavity $Q$-factor and the magnitude and sign of the higher-order nonlinear coefficients ($n_2/n_4$ ratio). These results demonstrate that higher-order nonlinearities must be considered when predicting the performance of resonant photonic structures operating under high circulating intensities. The results presented here therefore demonstrate how higher-order nonlinear effects can influence cavity-enhanced photonic devices when moving towards tighter confinement cavities, and when quality factors of such cavities continue to increase. Our results demonstrate how vertical Bragg microcavities could act as efficient nonlinear devices and as an interesting platform to study strongly nonlinear light–matter interactions.

\begin{acknowledgments}
We wish to acknowledge the support of Finnish Foundation for Technology Promotion (Grant No. 10886) and the Flagship of Photonics Research and Innovation (PREIN) funded by the Academy of Finland (Grant No. 320165).
\end{acknowledgments}

\section*{Data availability}
The data that support the findings of this study are publicly available.

\section*{Supplemental Material}
This supplemental material provides additional details on the higher-order Kerr effect including both positive and negative values of $n_4$, the optical bistability calculations and hysteresis curves, the curved-cavity simulations and radius-of-curvature calculations, and the $Q$-factor analysis of high-$Q$ cavities.

\bibliography{apssamp.bib}

\end{document}


\renewcommand{\figurename}{Fig.}
\renewcommand{\thefigure}{S\arabic{figure}}
\renewcommand{\theequation}{S\arabic{equation}}

\title{\textbf{Vertical microcavities with optical Kerr materials} 
}%


\author{Riya Varghese}
\email[]{riya.varghese@tuni.fi}
\author{Ali Panahpour} 
\email[]{ali.panahpour@tuni.fi}
\author{Marco Ornigotti} 
\email[]{marco.ornigotti@tuni.fi}
\author{Mikko J Huttunen}
\email[]{mikko.huttunen@tuni.fi}
\affiliation{Photonics Laboratory, Physics Unit, Tampere University, Korkeakoulunkatu 3, 33720 Tampere, Finland}


\date{\today}


\maketitle
\subsection{Higher-order Kerr Effect}
\begin{figure}
\includegraphics{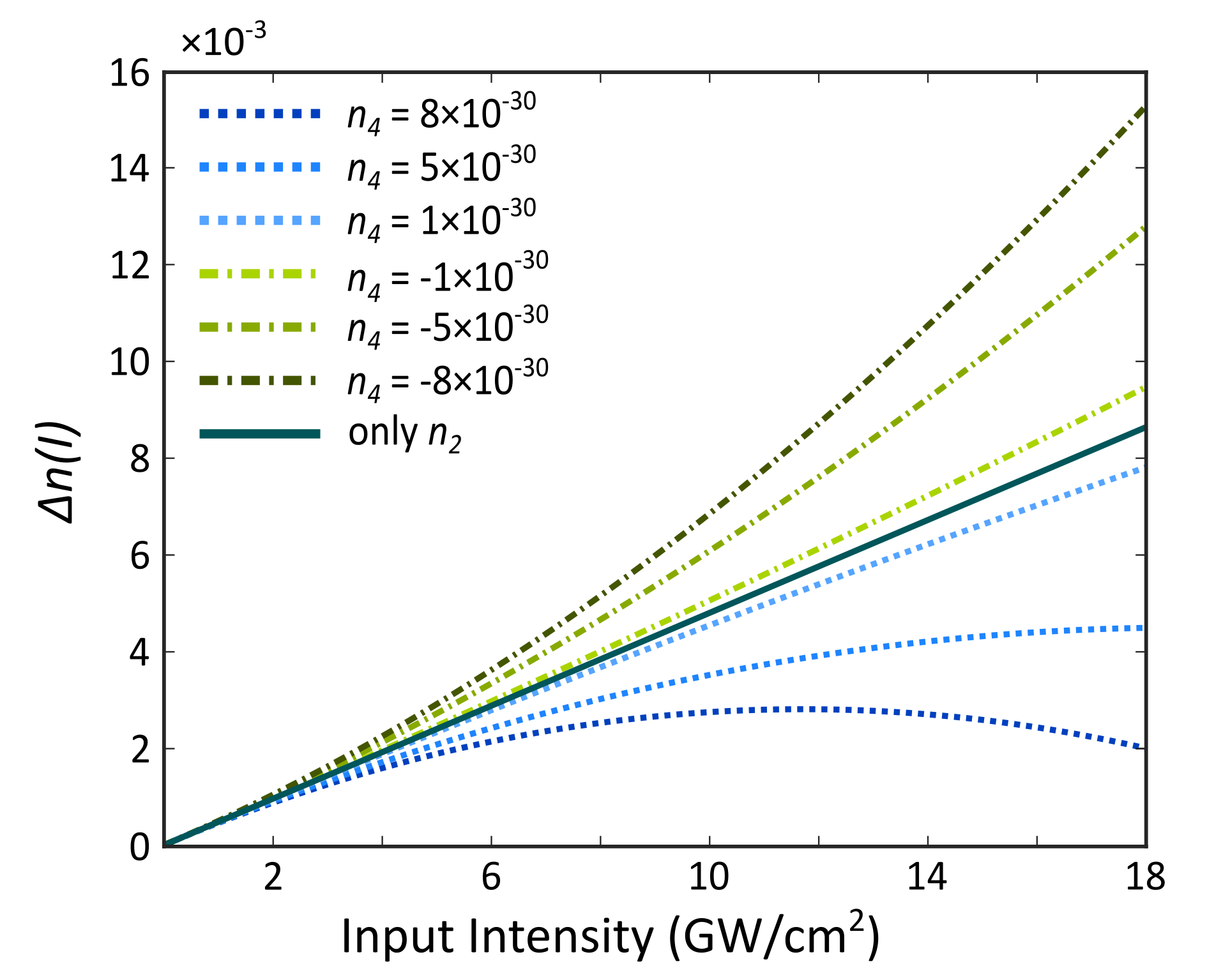}
\caption{The refractive index shift ($\Delta n$) as a function of optical intensity for positive and negative values of the fourth-order nonlinear coefficient $n_4$.}
\label{fig:S1}
\end{figure} 
\noindent As seen in Eq.~(4) in main document, the coefficient $n_4 n_0/n_2^2$ determines how the refractive index changes with respect to input intensity at higher intensities where higher-order Kerr effect (HOKE) becomes relevant. To illustrate the effect of the sign of the nonlinear coefficient, we compare the refractive index change obtained for positive and negative values of the fourth-order nonlinear refractive index coefficient ($n_4$). As shown in supplementary Fig.~\ref{fig:S1}, positive $n_4$ enhances the refractive index shift, whereas negative $n_4$ reduces the overall index change and can lead to saturation of the nonlinear response at high intensities. Both positive and negative values of $n_4$ have been reported in the literature depending on the medium, wavelength, and intensity regime. For instance, calculations in noble gases (He, Ne, Ar, Xe) shows a positive value of  $n_4$ in the non-resonant regime \cite{Tarazkar2014High-orderAtoms}. In contrast, semiconductors and filamentation in gases have reported effective negative higher-order nonlinearities \cite{Chen2006MeasurementGlasses}, resulting in saturation of the refractive index change at high intensities.

\subsection{Optical Bistability}
\begin{figure}
\includegraphics{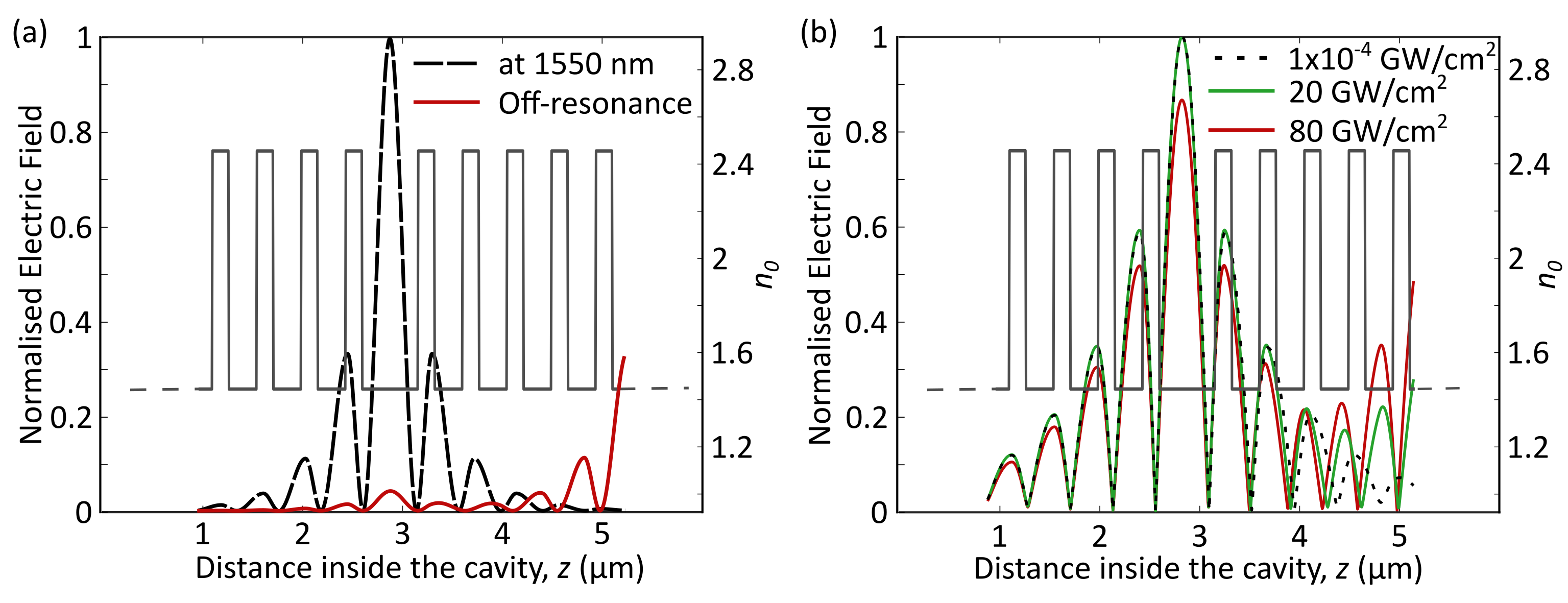}
\caption{a) The field inside the Bragg cavity at the resonant wavelength and at off-resonant wavelength without the Kerr effect. On resonance, the $E$-field is maximized inside the core region, decaying rapidly within the Bragg mirror regions. At off-resonant wavelengths, the $E$-field is not strongly confined within the core region. b) The field inside the cavity at the designed wavelength (1550 nm) when a Kerr-induced refractive index change is introduced in the core. As the input intensity increases, the cavity resonance red-shifts away from the designed wavelength, resulting in progressively weaker field confinement within the core region.}
\label{fig:S2}
\end{figure}

\noindent The optical bistability is modeled using the transfer-matrix-method implemented in MATLAB which is used to determine the intracavity intensity and field-enhancement (Fig.~\ref{fig:S2}a) for each input intensity for the designed Bragg cavity. The material dispersion was included in the calculations using the Sellmeier equation for both SiO$_2$ \cite{Malitson1965InterspecimenSilica} and TiO$_2$ \cite{DeVore1951RefractiveSphalerite}. For a given input intensity, the intracavity field at the designed wavelength is first calculated (Fig.~\ref{fig:S2}b) and used to evaluate the Kerr-induced modification of the refractive index according to Eq.~(3) in main document. The cavity response is subsequently recomputed using the updated refractive index profile. These steps are iterated within a self-consistent loop until convergence is reached. Specifically, until the maximum change in the field between successive iterations was below 0.001 yielding a stable field distribution and refractive index profile for the selected input intensity.

To model the experimentally observable transmission (Fig~4d--f in main document), the cavity response was coupled to an external excitation source centered at 1550 nm. The cavity linewidth was approximately 2~nm for a quality factor of $Q$\;=\;1000. An excitation source with bandwidth of 4~nm was used, chosen to maximize the overlap between the excitation source and the cavity resonance. As the input intensity increases, the Kerr-induced refractive index change shifts the cavity resonance away from the excitation wavelength. Consequently, the source-cavity overlap decreases, giving rise to the decrease in transmission response and eventually optical bistability. The corresponding field distribution inside the cavity at the designed wavelength is shown in Fig.~\ref{fig:S2}b. At the lowest input intensity ($I_{in} = 1\times10^{-4}$ GW/cm$^2$), the field is strongly confined within the core region, indicating that the cavity remains resonant at the designed wavelength. As the input intensity increases to $I_{in}$ = 80 GW/cm$^2$, the Kerr-induced refractive index change causes the cavity resonance to red-shift away from the designed wavelength. Consequently the field is no longer strongly localized in the core region.

The hysteresis curve in Fig.~4c in main document was obtained by performing both forward and backward intensity sweep for a $Q$-factor of 1000 and at a wavelength of 1550~nm. For the forward sweep, the input intensity was gradually increased, and the intracavity intensity was calculated at each step to find the stable field distribution. The same procedure was repeated while decreasing the input intensity to generate the reverse branch of the hysteresis loop. The difference between the forward and reverse sweeps produces the characteristic bistable response. Initially, the laser wavelength is at the cavity resonance which makes the light couple into the cavity efficiently. As we increase the input intensity, the cavity redshifts transmission decreases (and reflection correspondingly increases) as the resonance shifts away from the excitation wavelength. As a result, the intracavity field further decreases (Fig.~\ref{fig:S2}b) and changes the nonlinear refractive index shift, creating a positive feedback mechanism that leads to the characteristic hysteresis behaviour of optical bistability.

\subsection{Curved Cavities}
\begin{figure}
\includegraphics{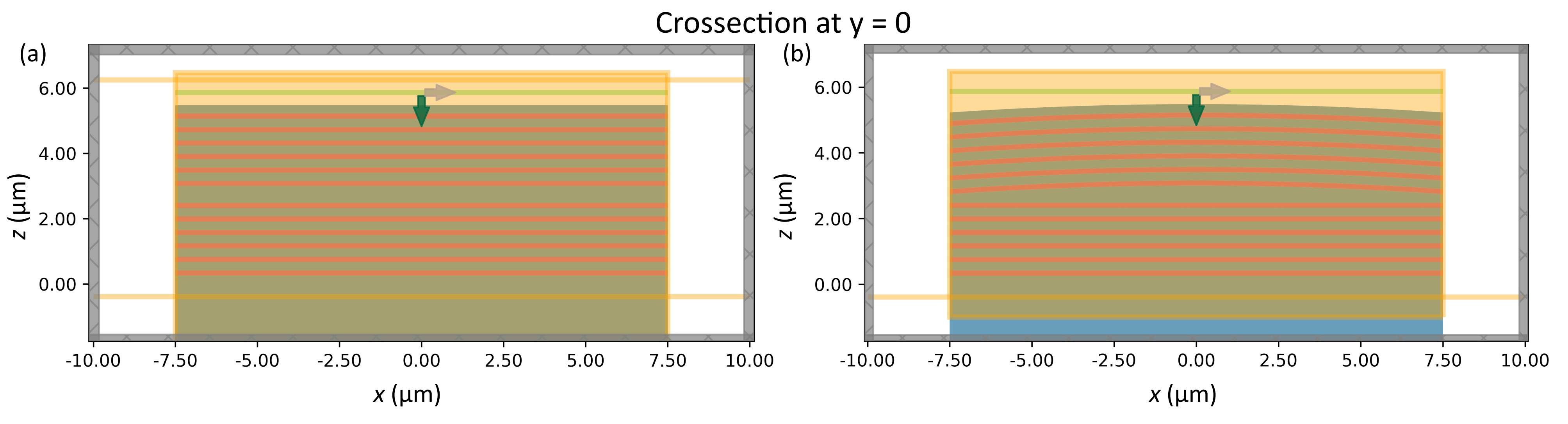}
\caption{a) The simulation domain of the flat Bragg cavity structure with the source and monitors. b) The simulation domain of the curved Bragg cavity structure with the source and monitors.}
\label{fig:S3}
\end{figure} 
\noindent The simulations of finite curved Bragg cavities were performed using the FDTD solver in Tidy3D. The structures were modelled as two-dimensional ridge cavities to reduce the computational cost. Periodic boundary conditions were used along the lateral $y$-axis and perfectly matched layer boundary conditions were applied along the propagation direction and vertical direction ($z$- and $x$-axes respectively). A Gaussian beam source was launched into the cavity and the transmission response was monitored using flux monitors. Unlike planar cavities, the resonance condition of curved cavities depends on Gaussian beam optics and the position of beam waist. The Gaussian beam source was defined by a beam waist ($w_0$) of 2~$\mu$m, positioned at the center of the cavity core, from which the corresponding Rayleigh range is calculated as
\begin{equation}
z_R = \frac{\pi w_o^2 n}{\lambda}\;,
\label{eq:rayleigh range}
\end{equation}
where $n$ is the index of the core of the cavity.  For a given cavity length ($d_c\approx\frac{\lambda}{2n}$), the remaining Gaussian beam parameters, including the Rayleigh range ($z_R$) and the wavefront radius of curvature  
\begin{equation}
R(z) = z\left[1 + \left(\frac{z_R}{z}\right)^2\right]\;, 
\label{eq:Radius}
\end{equation}
were calculated externally in MATLAB. The influence of the Gouy phase on the cavity resonance was accounted for through the resonance condition reported in the literature \cite{Koks2021MicrocavityDepths}. The resonance was identified by keeping the cavity length fixed and varying the radius of curvature ($R(z)$) in the simulation.

For cavities with moderate quality factors ($Q < 5000$), the resonance wavelength and quality factor could be extracted directly from the transmission spectrum
\begin{equation}
Q = \frac{\lambda_0}{\Delta \lambda}\;,
\label{eq:Q factor}
\end{equation}
where the $\lambda_0$ is the resonant wavelength and $\Delta \lambda$ is the full-width half maximum of the cavity resonance. However, as the cavity quality factor increases, leading to very long photon lifetimes \cite{Mandelshtam1997HarmonicApplications}, significantly longer simulation times are required for the electromagnetic fields to fully decay. In many cases, the required simulation duration exceeded the practical runtime limits of Tidy3D and resulted in prohibitively high computational cost.
To overcome these limitations, high-$Q$ cavity simulations were performed using the Tidy3D ResonanceFinder plugin \cite{ExampleFlexcompute}. Instead of exciting the cavity with a Gaussian beam and extracting resonance properties from a transmission spectrum, a dipole source was placed inside the cavity to excite its resonant modes over a broad spectral range. A flux time monitor was used to record the temporal decay of the electromagnetic field following excitation. The ResonanceFinder algorithm then analyzed the decay signal and directly extracted the resonance frequency, decay rate, and $Q$-factor of the cavity mode. The $Q$-factors were analyzed using both the transmission-based method and the ResonanceFinder approach for low-$Q$ cavities (number of pairs, $N$ = 4 and 6) giving  good agreement with each other. The Kerr effect in flat finite cavities were incorporated into the simulations using the nonlinear material model available in the Tidy3D framework.

\bibliography{apssamp.bib}